\documentclass{aa}  
\usepackage{graphicx}
\usepackage{txfonts}
\usepackage{xcolor}
\usepackage[switch]{lineno} 
\usepackage{amsmath}

\newcommand{\gsim}{\lower.5ex\hbox{$\; \buildrel > \over \sim \;$}}
\newcommand{\lsim}{\lower.5ex\hbox{$\; \buildrel < \over \sim \;$}}

\newcommand{\nv}{\hbox{N\,{\sc v}}}

\newcommand{\civ}{\hbox{C\,{\sc iv}}}
\newcommand{\cv}{\hbox{C\,{\sc v}}}
\newcommand{\siiv}{\hbox{Si\,{\sc iv}}}

\newcommand{\ciii}{\hbox{C\,{\sc iii}}}
\newcommand{\ovi}{\hbox{O\,{\sc vi}}}

\newcommand{\kms}{\ifmmode {\rm km\,s}^{-1} \else km\,s$^{-1}$ \fi}
\newcommand{\cc}{\hbox{cm$^{-3}$}}

\newcommand{\ergcms}{\ifmmode {\rm ergs\,cm}^{-2}\,{\rm s}^{-1} \else ergs\,cm$^{-2}$\,s$^{-1}$\fi}
\newcommand{\ergcmsA}{\ifmmode{\rm ergs}\, {\rm cm}^{-2}\,{\rm s}^{-1}\,{\rm\AA}^{-1} \else ergs\, cm$^{-2}$\, s$^{-1}$\, \AA$^{-1}$\fi}
\newcommand{\ergcmsHz}{\ifmmode{\rm ergs\,cm}^{-2}\,{\rm s}^{-1}\,{\rm Hz}^{-1} \else ergs\,cm$^{-2}$\,s$^{-1}$\,Hz$^{-1}$\fi}
\newcommand{\phcms}{\ifmmode {\rm ph\,cm}^{-2}\,{\rm s}^{-1} \else ,ph\,cm$^{-2}$\,s$^{-1}$\fi}
\newcommand{\phcmsA}{\ifmmode {\rm ph\,cm}^{-2}\,{\rm s}^{-1}\,{\rm\AA}^{-1} \else ph\,cm$^{-2}$\,s$^{-1}$\,\AA$^{-1}$\fi}

\newcommand\Msun{\ifmmode M_{\odot} \else $M_{\odot}$\fi}
\newcommand\msun{\ifmmode M_{\odot} \else $M_{\odot}$\fi}
\newcommand\Lsun{\ifmmode L_{\odot} \else $L_{\odot}$\fi}
\newcommand\Zsun{\ifmmode Z_{\odot} \else $Z_{\odot}$\fi}
\newcommand\mpyr{\ifmmode \Msun\,{\rm yr}^{-1} \else $\Msun\,{\rm yr}^{-1}$ \fi}

\newcommand{\Luv}{\ifmmode L_{1450} \else $L_{1450}$\fi}
\newcommand{\Lop}{\ifmmode L_{5100} \else $L_{5100}$\fi}
\newcommand{\Lthree}{\ifmmode L_{3000} \else $L_{3000}$\fi}
\newcommand{\lledd}{\ifmmode L/L_{\rm Edd} \else $L/L_{\rm Edd}$\fi}
\newcommand{\ledd}{\ifmmode L_{\rm Edd} \else $L_{\rm Edd}$\fi}
\newcommand{\lamLlam}{\ifmmode \lambda L_{\lambda} \else $\lambda L_{\lambda}$\fi}
\newcommand{\lbol} {\ifmmode L_{\rm bol} \else $L_{\rm bol}$\fi}
\newcommand{\llbol}{\ifmmode \log\left(\lbol/\ergs\right) \else $\log\left(\lbol/\ergs\right)$\fi}

\newcommand{\fuv}{\ifmmode f_{\lambda}\left(1450\AA\right) \else $f_{\lambda}\left(1450 {\rm \AA}\right)$\fi}
\newcommand{\fthree}{\ifmmode f_{\lambda}\left(3000\AA\right) \else $f_{\lambda}\left(3000{\rm \AA}\right)$\fi}
\newcommand{\fH}{\ifmmode f_{\lambda}\left(1.65\micron\right) \else
$f_{\lambda}\left(1.65\micron\right)$\fi}

\newcommand{\mbh}{\ifmmode M_{\rm BH} \else $M_{\rm BH}$\fi}
\newcommand{\lmbh}{\ifmmode \log\left(\mbh/\Msun\right) \else $\log\left(\mbh/\Msun\right)$\fi}

\def\dif{\mathop{}\hphantom{\mskip-\thinmuskip}\mathrm{d}}%
\let\daccent\d
\gdef\d{\ifmmode\dif\else\expandafter\daccent\fi}

\begin{document}

   \title{Discovery of the Asymmetric Effect in the Response of Photoionization Gas}

   \author{Zhicheng He
          \inst{1,2}\fnmsep\thanks{zcho@ustc.edu.cn}
          \and
          Tinggui Wang \inst{1,2}
          }

   \institute{Department of Astronomy, University of Science and Technology of China, Hefei 230026, China\\
         \and
             School of Astronomy and Space Science, University of Science and Technology of China, Hefei, Anhui 230026, China\\
             }

   \date{Received XXX; accepted XXX}

  \abstract
   {Ionized gas is ubiquitous in the universe and plays a central role in tracing cosmic evolution and probing plasma physics under extreme conditions. Among various ionizing sources, quasars—powered by supermassive black holes—are important contributors to the reionization of the universe. The variability of quasar radiation provides a valuable opportunity to study the photoionization response of interstellar and intergalactic gas.}
   {We aim to investigate the physical origin of the asymmetric response of ionized gas to variable quasar radiation, particularly as observed in broad absorption line (BAL) systems. We also seek to place constraints on the gas density and spatial scale of the BAL outflows based on this asymmetry.}
   {We conduct time-dependent photoionization simulations focusing on \civ\ to quantify the response timescales across different ionization states. Analytical estimates are also used to relate response asymmetry to gas density.}
   {We find that over 70\% of BAL gas in quasar host galaxies exhibit a negative response to quasar dimming, indicating a strong asymmetry in the ionized gas behavior. Our simulations show that this asymmetry arises from shorter response timescales at higher ionization states. Given typical observational cadences (>1 day), the observed asymmetry requires at least 40\% of the BAL gas to have a density below $n_{\rm H} = 10^6\ \cc$, consistent with most measured BAL gas densities. This is in contrast to the typical density of accretion disk winds ($n_{\rm H} > 10^8\ \cc$), suggesting that BAL outflows either evolve significantly as they propagate outward or may originate from larger-scale regions such as the dusty torus.}
   {We have uncovered a fundamental asymmetry in the response of ionized gas: the response timescales of high-ionization states are shorter than those of low-ionization states.
The role of asymmetric response effects thus offers new constraints on the physical origin and structure of quasar outflows.}
   \maketitle

\section{Introduction}
\label{sec:introduction}
Ionized gas is prevalent in the universe, at high redshifts ($z>1100$) and after the cosmic reionization epoch with $z<6$ 
\citep{barkana2001,kriss2001,zheng2004,fan2006,tilvi2020,yung2020}. Astrophysical plasma, found in entities such as the interstellar medium (ISM) and intergalactic medium (IGM), 
serves as a valuable tool for exploring the evolutionary processes of both the universe and celestial bodies. 
Simultaneously, cosmic celestial bodies provide an ideal laboratory for studying the principles of plasma physics in extreme situations such as extremely high temperatures or 
low densities. Quasars, a class of high luminosity active galactic nuclei (AGN), fueled by the accretion disk of supermassive black holes (SMBHs), stand 
out as the most luminous and persistent celestial objects in the universe, with 
their redshift frontier extending to approximately z $\sim$ 7.6 \citep{wang2021}. Quasars is believed to play crucial roles in both cosmic reionization \citep{jakobsen1994,loeb2001,fan2023} 
and the evolution of galaxies \citep{fabian2012,kormendy2013,veilleux2020},
through the intense ionizing radiation and enormous energy released outward. 

\begin{figure}[htb]
\centering
\includegraphics[width=0.5\textwidth]{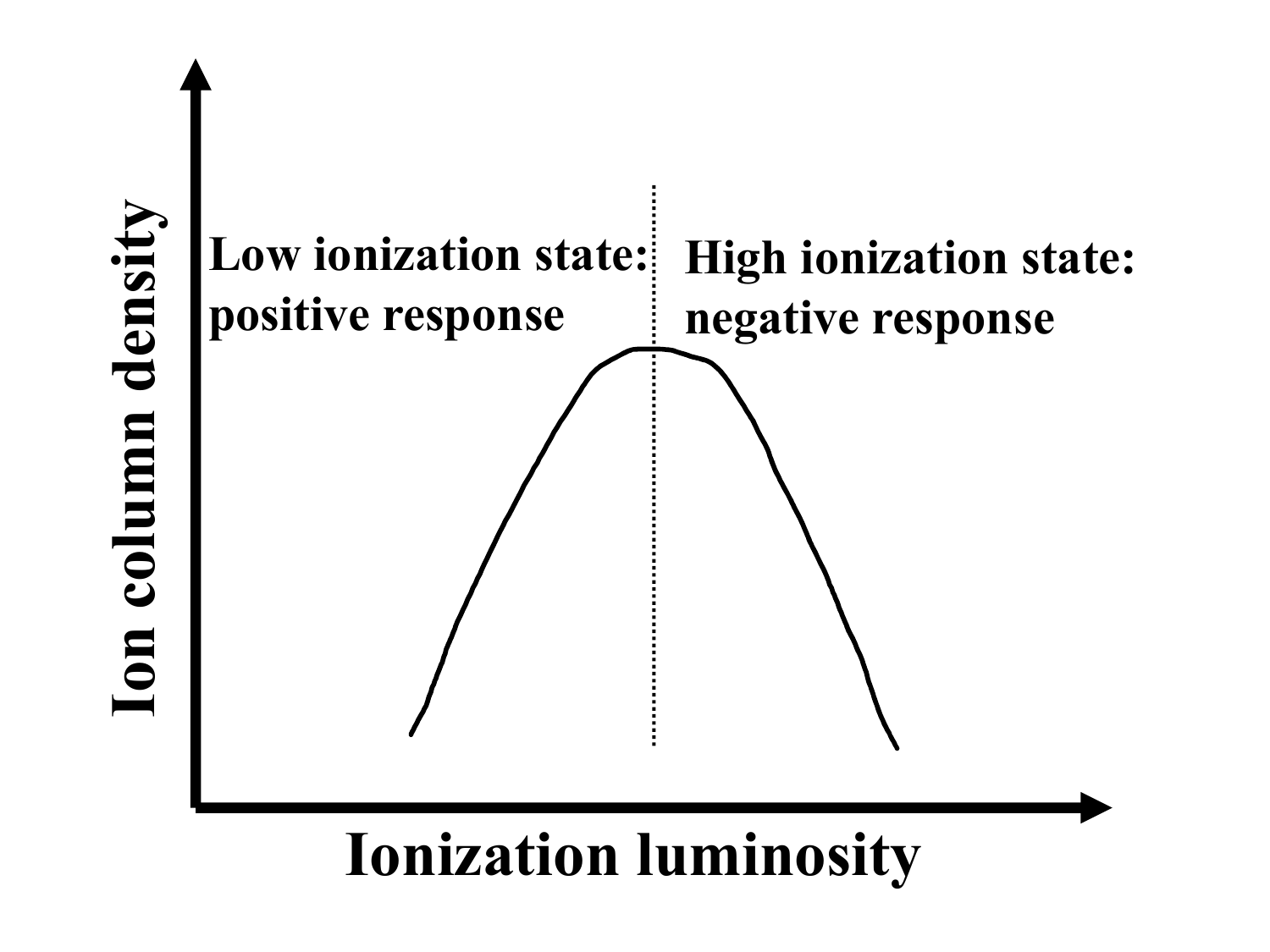}
\caption{Schematic diagram of the variation of ion column density with ionization luminosity.
The column density of a specific ion in the plasma initially increases to a peak and then decreases with the rise in ionization level. 
The vertical dashed line represents the peak position.
Hence, based on the peak position, it can be classified into two stages: the low ionization state and the high ionization state.
\label{fig1}}
\end{figure}

The radiation of quasar typically displays aperiodic variabilities, often described by the damped random walk (DRW) model \citep{kelly2009}. 
The variability of quasar radiation provides a useful probe of the response of ionized gas, as well as its density and spatial distribution \citep{nicastro1999,krongold2007,kaastra2012,garcia2013,he2014,he2015,he2019,he2022,rogantini2022,sadaula2023,sadaula2024}.
The ionization state of a gaseous outflow needs a period of time to respond to changes in the ionizing continuum [the recombination timescale $t^*$, e.g, \cite{krolik1995}]. 
Therefore, the gas ionization state is connected to the intensity of the ionizing continuum over $t^*$.
At this point, the function of the $t^*$ is to induce a smoothing effect.  As illustrated in Fig. \ref{fig1}, the population density of a 
specific ion in the plasma initially increases to a peak and then decreases with the rise in ionization level. Hence, based on the peak position, it can be classified into two 
stages: the low ionization state and the high ionization state. 
In the low-ionization state, the ion column density increases with rising ionizing luminosity, indicating a positive correlation between the two, referred to as a positive response. 
Conversely, in the high-ionization state, the ion column density decreases as the ionizing luminosity increases, reflecting a negative correlation between the two, known as a negative response.

Some studies (e.g., \citealt{wang2015,he2017}) utilizing the large sample from  
Sloan Digital Sky Survey (SDSS) have revealed that over 70\% of \siiv\ (ionization energy 45.1 ev), \civ\ (64.5 ev), \nv\ (97.9 ev) broad absorption lines (BALs) 
exhibit a negative response. In other words, the majority of gases with variable BALs are in a highly ionized state. Naturally, there are some speculations regarding this.
Does this negative response persist for ions with higher ionization energies, such as \ovi\ (138.1 ev), or highly ionized iron (>1 kev) ?
Is it because most BAL gases are in a highly ionized state, or is there some physical asymmetry in the plasma that makes highly ionized ions more prone to 
responding to variations in the radiation source?

In this study, we explored the asymmetry of plasma response to radiation sources for low and high ionization states. Section 2 presents a comprehensive analysis of the 
asymmetry in plasma response time scales. In Section 3, we apply this asymmetry to \civ\ using photoionization simulation. The final section provides a summary.

\section{The asymmetric effect of response timescale}
Studies of BALs in quasar spectra have shown that ions such as $\rm N^{4+}$, $\rm C^{3+}$, and $\rm Si^{3+}$ in the gas surrounding quasars are 
primarily governed by photoionization, e.g., \cite{wang2015,he2017,he2022,lu2018,hemler2019,vivek2019,zhao2021}.
Therefore, our work focuses on gases dominated by photoionization.
Here, we will provide an overview of the derivation process for the precise recombination timescale of the photoionization gas in previous literature (e.g., \citealt{arav2012, netzer2013}).
The population density of of a given element in ionization stage i is represented by 
\begin{eqnarray}  \label{eq1}
\frac{\d n_i}{\d t}=-n_i(I_i+R_{i-1})+n_{i-1}I_{i-1}+n_{i+1}R_{i},
\end{eqnarray}
where the ionization rate per particle is $I_i$, and the recombination rate per particle from ionization stage i+1 to i is 
$R_i$. 
Here, we exclusively focused on the photoionization processes and omitted collisional ionization from our analysis.
Equation \ref{eq1} forms a set of n+1 coupled differential equations for an element with n electrons and n+1 ions.
In the state of photoionization equilibrium, i.e., when ${\d n_i}/{\d t}=0$, these reduce to n equations of the form
\begin{eqnarray}  \label{eq2}
\frac{n_{i+1}}{n_{i}}=\frac{I_{i}}{R_{i}}.
\end{eqnarray}
This implies that the increase of stage i by recombination from stage i+1 must be balanced by the decrease of stage i by ionization to stage i+1.
For an optically thin gas at distance $r$ from the radiation source with spectral luminosity $L_{\nu}$ at frequency $\nu$, the ionization rate ($\rm s^{-1}$) 
per ion of stage i is given by
\begin{eqnarray}  \label{eq3}
I_i=  \frac{\Gamma_i }{  {4\pi r ^2} } ,
\end{eqnarray}
$\Gamma_i$ is the integral ionization cross-section rate ($\rm cm^{2} \ s^{-1}$)  over the 
photons with frequency $\nu > \nu_i$ :
\begin{eqnarray}  \label{eq4}
\Gamma_i=\int_{\nu_i}^{\infty} (L_{\nu}/h\nu) \sigma_{\nu}   \d\nu ,
\end{eqnarray}
where $h$ is Planck's constant and $\sigma_{\nu}$ is the ionization cross-section for photons of energy $h\nu$. The
$\sigma_{\nu}$ for different atoms and ions can be obtained from the fitting formula in Equation 1 and Table 1 in \citep{verner1996}.
The recombination rate per particle is given by 
\begin{eqnarray}  \label{eq5}
R_i= \alpha_i(T)n_e,
\end{eqnarray}
where the recombination coefficient $\alpha_i$ depends on the electron 
temperature $T$ \citep{osterbrock2006}.

\begin{figure}[htb]
\centering
\includegraphics[width=0.5\textwidth]{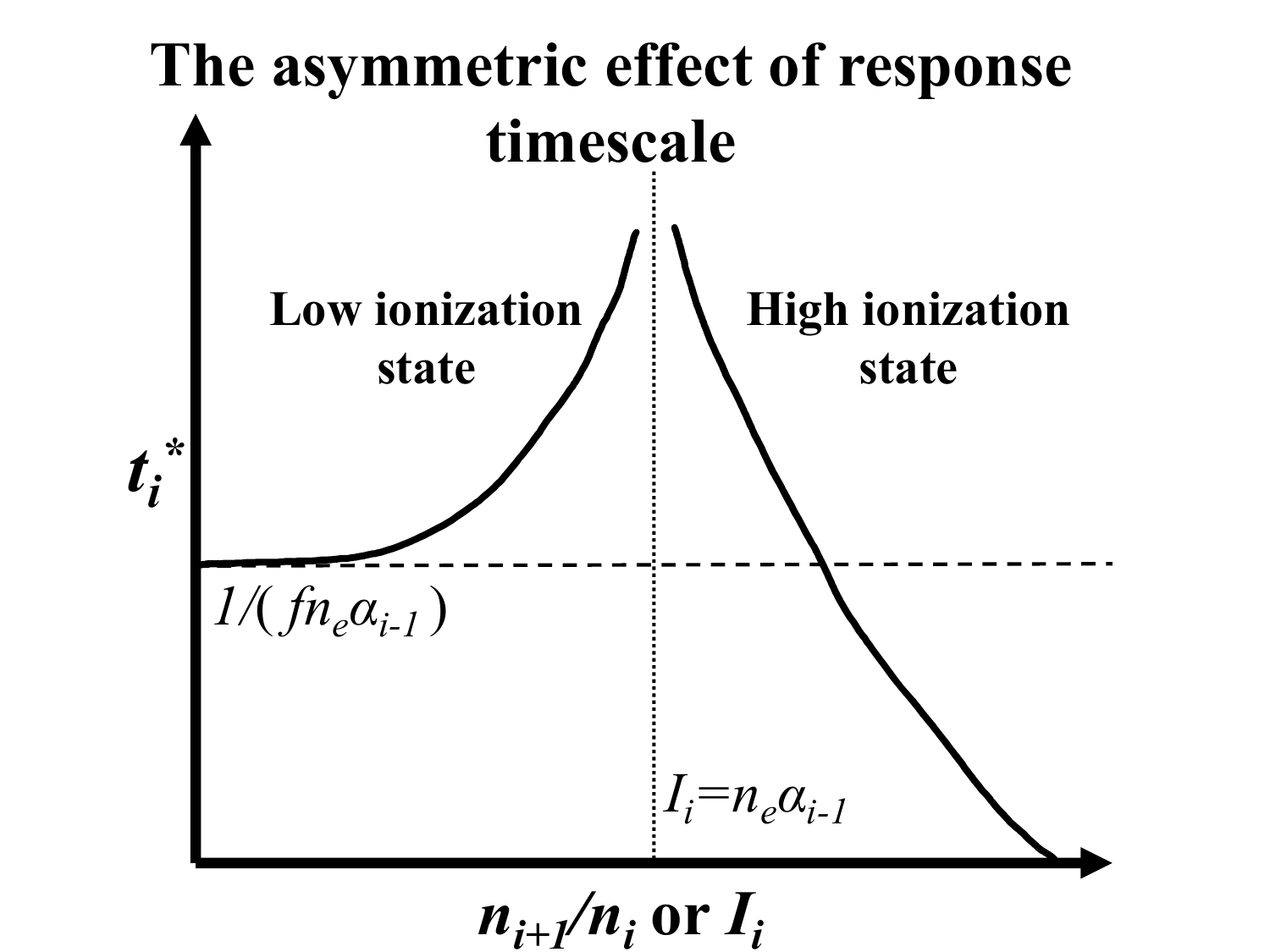}
\caption{Schematic diagram of the asymmetric effect of ion response timescale.
The vertical dashed line is the boundary between the low ionized state and the high ionized state.
In a sufficiently low ionization state, i.e., $n_{i+1}/n_{i} \ll \alpha_{i-1}/\alpha_{i}$ or $I_{i} \ll n_e\alpha_{i-1}$,
the response timescale is $t_i^*=1/(fn_e\alpha_{i-1})$. On the contrary, in a sufficiently high ionization state, 
i.e., $n_{i+1}/n_{i} \gg \alpha_{i-1}/\alpha_{i}$ or $I_{i} \gg n_e\alpha_{i-1}$,
the response timescale is $t_i^*=-1/(f\alpha_{i} n_e \frac{n_{i+1}}{n_{i}})=-1/(fI_i)$. 
In highly ionized state, the response timescale is shorter than that of low ionized state, leading to the detection of more 
negative responses.
\label{fig2}}
\end{figure}

Suppose an absorber in photoionization equilibrium experiences a sudden change in the incident ionizing flux such that $I_i(t > 0)=(1+f)I_i(t=0)$, 
where $-1\le f \le +\infty$. 
At this moment, the column densities of adjacent valence ions and recombination coefficients have not yet had the opportunity to adjust in response to the 
sudden change of the incident photon flux, i.e., $\d n_i$ = $\d n_{i-1}$ = $\d n_{i+1}$ = $\d R_i$ = $\d R_{i-1}$ = 0. 
Therefore, based on Equation \ref{eq1} and \ref{eq2}, the evolution of $n_{i}$ can be written as:
\begin{eqnarray}  \label{eq6}
\frac{\d n_i}{\d t}=-fn_iI_i+fn_{i-1}I_{i-1}= -fn_iI_i+fn_{i}R_{i-1}.
\end{eqnarray}
Then the response timescale for change in the ionic fraction is:
\begin{eqnarray} 
t_i^*=\frac{\d t}{\d \ln n_i} =&\left[f \alpha_{i} n_{e}\left(\frac{\alpha_{i-1}}{\alpha_{i}}-\frac{n_{i+1}}{n_{i}}\right)\right]^{-1},\ {\rm or} \label{eq7}\\
 = &\left[f (n_{e}\alpha_{i-1}-I_i)\right]^{-1}. \label{eq8}
\end{eqnarray}
The response time scale here is the e-folding time, i.e., the time required for the initial column density to increase e times or decrease to 1/e based on the initial rate of change. The e-folding time is directly proportional to the equilibrium time, with only one additional factor $f$ compared to the equilibrium time.
Strictly speaking, the above equations are valid only at the initial moment. Nevertheless, they remain highly informative, as they reveal both the sign of the response 
(positive or negative) and the response timescale over which the ion concentration evolves.
As shown in Fig.~\ref{fig2}, the regime can be divided into high- and low-ionization states according to the position of 
$n_{i+1}/n_{i} =\alpha_{i-1}/\alpha_{i}$ or $I_{i} = n_e\alpha_{i-1}$. 
It is worth noting that there is a misalignment between the boundary between the low- and high-ionization states and the boundary defined by the peak column density. 
This misalignment gives rise to a novel mixed-response effect \citep[see][]{he2025}. 
However, this effect arises only near the intermediate boundary and does not affect the asymmetry between the extremely low- and extremely high-ionization 
states that are the focus of this work.
When the gas is in a sufficiently low ionization state, i.e., $n_{i+1}/n_{i} \ll \alpha_{i-1}/\alpha_{i}$ or $I_{i} \ll n_e\alpha_{i-1}$,
the response timescale is $t_i^*=1/(fn_e\alpha_{i-1})$. On the contrary, when the gas is in a sufficiently high ionization state, 
i.e., $n_{i+1}/n_{i} \gg \alpha_{i-1}/\alpha_{i}$ or $I_{i} \gg n_e\alpha_{i-1}$,
the response timescale is $t_i^*=-1/(f\alpha_{i} n_e \frac{n_{i+1}}{n_{i}})=-1/(fI_i)$ = $-r^2/(f\Gamma_i)$. 
Clearly, this timescale exhibits asymmetry for low ionization and high ionization states.
In the low ionization state, the response timescale is dominated by the recombination coefficient (i.e., the so-called recombination timescale), 
whereas in the high ionization state, it is dominated by the ionization rate. 
In a highly ionized state, the $t^*$ is shorter, that is, when the radiation source flux changes, the time required to return to 
gas ionization equilibrium is shorter. 
The preceding discussion is based on theoretical analysis. In the next section, we will test these conclusions 
through time-dependent photoionization simulations.

\section{Photoionization simulations for carbon ions}
In this section, we will begin with outlining the detailed parameter settings taken in our photoionization simulations. 
Then, we use the ion $\rm C^{3+}$ as a case study to illustrate the asymmetric response.

\begin{figure}[htb]
\centering
\includegraphics[width=0.5\textwidth]{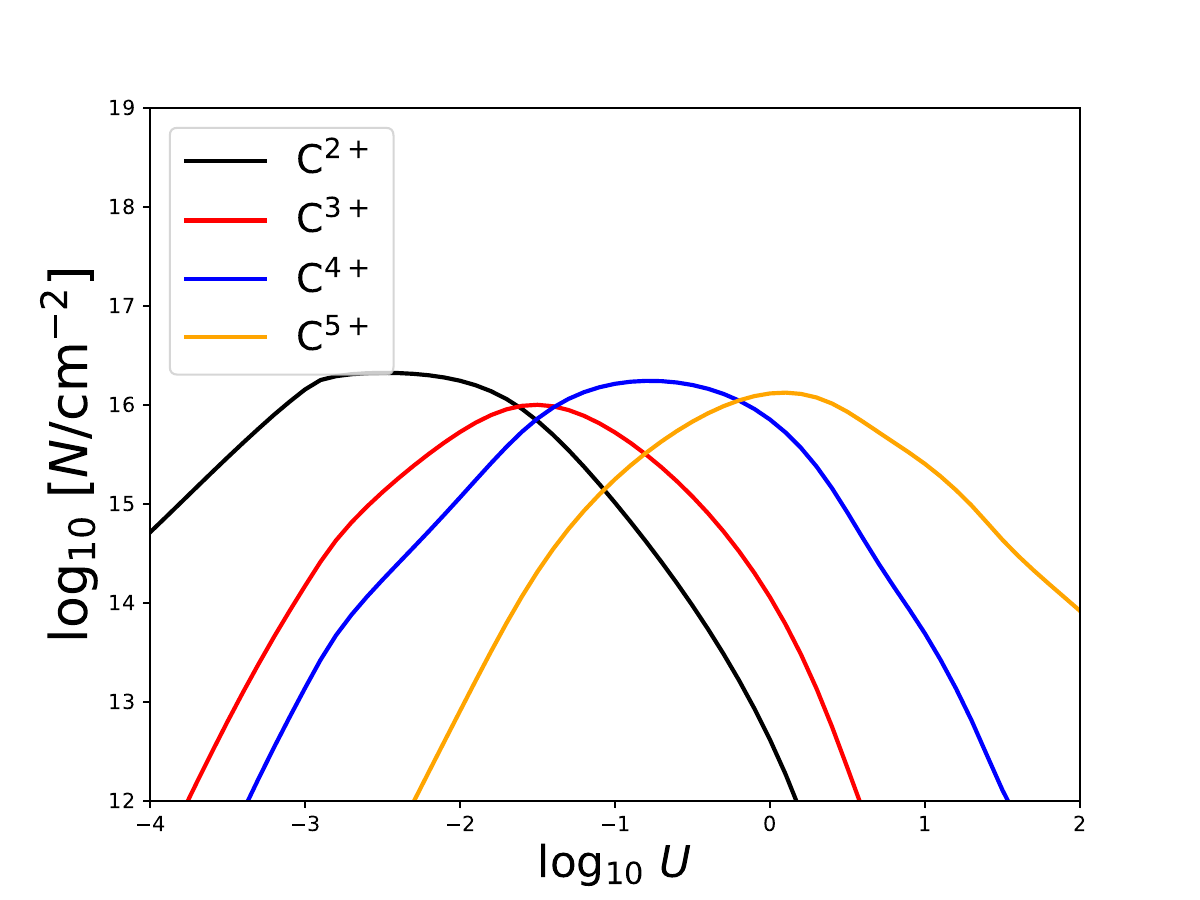}
\caption{Non-Monotonic Dependence of Carbon Ion Column Densities on Ionization Parameter.
The column densities of carbon ions increase first and then decrease with ionization parameters.
Based on the peak position of each ion, we can divide the parameter space into corresponding low- and high-ionization intervals. 
\label{fig3}}
\end{figure}

We adopt typical parameters for BAL gas in quasars, with two gas densities of $n_{\rm H} = 10^4,\  10^6 \cc$ and a column density of $N_{\rm H} = 10^{20}\ \rm cm^{-2}$.
Using CLOUDY simulations \citep{ferland2017}, we compute models across a wide ionization parameter range, $-4 < \log_{10} \ U < 2$, 
with a step size of $\Delta \log_{10} \ U = 0.1$. The spectral energy distribution (SED) of the ionizing radiation source is UV-SOFT \citep{dunn2010}, representative of 
high-luminosity, radio-quiet quasars. A standard solar metal- licity was adopted for the gas.
The ionization parameter is defined as follows:
\begin{eqnarray}\label{eq9}
U=\frac{Q_{\rm H}}{4\pi r^{2}n_{\rm H}c},
\end{eqnarray}
where
\begin{eqnarray}\label{eq10}
Q_{\rm H}=\int_{\nu_0}^{+\infty} \frac{L_{\nu}}{h\nu}\d \nu
\end{eqnarray}
is the source emission rate of hydrogen-ionizing photons and $h\nu_0$= 13.6 eV is the ionization potential of $H^0$ for ionization out of the 1-s ground state. 
$c$ is the speed of light in vacuum.
As shown in Fig.~\ref{fig3}, the column densities of carbon ions first increase and then decrease with the ionization parameter. 
Based on the peak position of each ion, we can divide the parameter space into corresponding low- and high-ionization intervals. 
Moreover, as the ionization energy of the ions increases, the transition boundary between the low- and high-ionization states 
systematically shifts toward higher ionization parameters.

\subsection{The asymmetric of response timescales of \civ}
\begin{figure}[htb]
\centering
\includegraphics[width=0.5\textwidth]{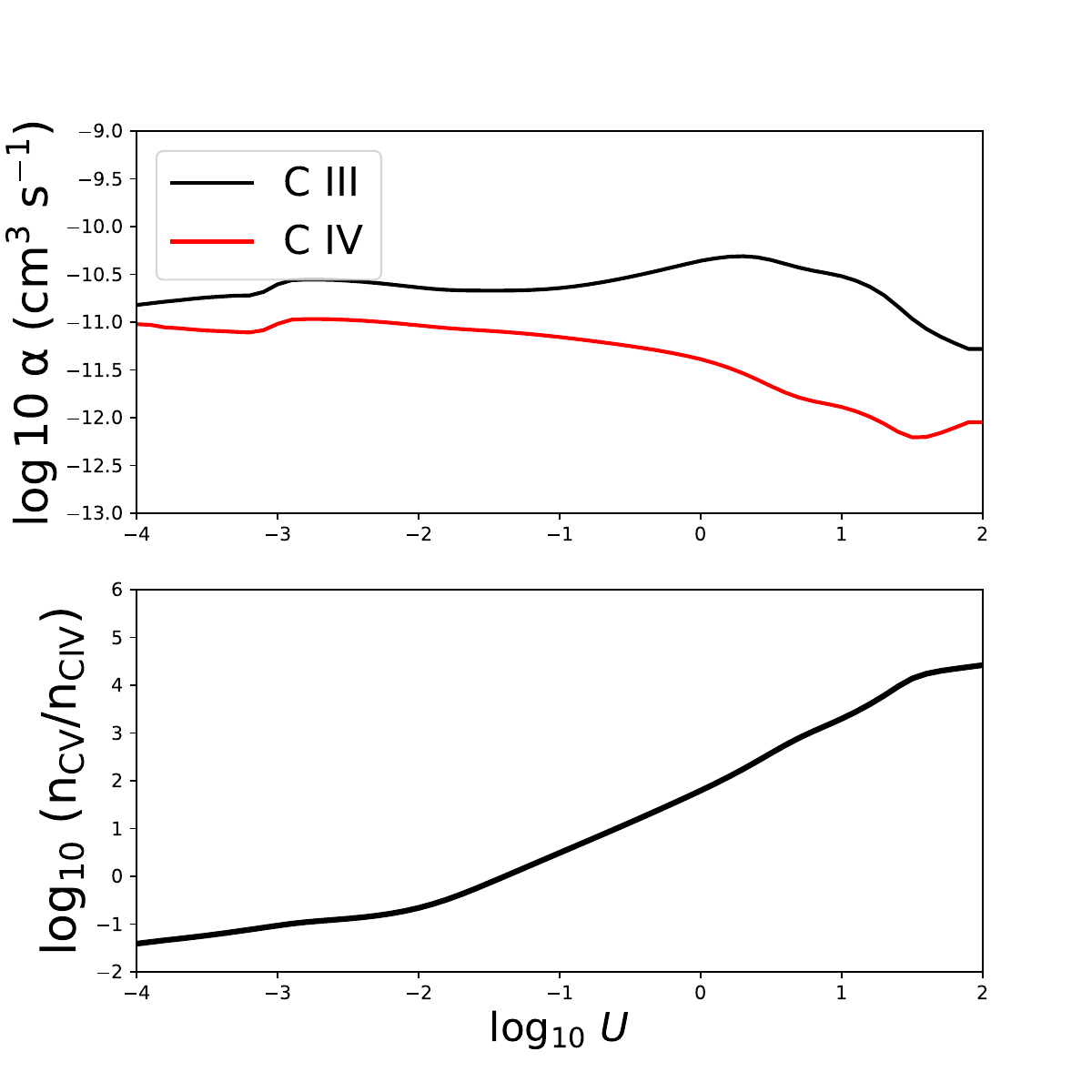}
\caption{The recombination coefficients and ion concentrations predicted by equilibrium-state photoionization simulations.
The recombination coefficients of \ciii\ and \civ, as well as the population density ratio $n_{\tiny \rm \cv}/n_{\tiny \rm \civ}$ under different ionization parameters,  
were obtained from equilibrium photoionization simulations using \textit{Cloudy} version c17 \citep{ferland2017}.
\label{fig4}}
\end{figure}

According the Equation \ref{eq7}, the expression for the response timescale of \civ\ (i.e., $\rm C^{3+}$) is
\begin{eqnarray}  \label{eq11}
t^*_{\tiny \civ}= \left[f \alpha_{\tiny \rm \civ} n_{e}\left(\frac{\alpha_{\tiny  \rm \ciii}}{\alpha_{\tiny  \rm \civ}}-\frac{n_{\tiny  \rm \cv}}{n_{\tiny  \rm \civ}}\right)\right]^{-1}.
\end{eqnarray}
In our calculation, we adopt a typical variation amplitude $f=1$ in the ionizing radiation band of AGN (mainly extreme ultraviolet).
As shown in Fig. \ref{fig4}, the curves of recombination coefficients of \ciii\ and \civ\ and the ratio of population density$n_{\tiny  \rm \cv}/n_{\tiny  \rm \civ}$ 
as a function of ionization parameters $U$ can be obtained from the photoionization simulations using Cloudy c17 \citep{ferland2017}.
As shown in Fig. \ref{fig5}, we calculate the value of the $t^*(\civ)$ at $\log10 \ U$ from -4 to 2 for $n_{\rm H} =10^4\  \cc$ 
and $10^6\  \cc$, respectively, based on the curves of recombination coefficients of \ciii\ and \civ\ and the ratio of population density $n_{\tiny \rm \cv}/n_{\tiny \rm \civ}$. 
Consistent with theoretical expectations, high-ionization states exhibit markedly shorter response timescales compared to low-ionization states.
In the low ionization state, the $t^*(\civ)$ is about 100 and 1 days for $n_{\rm H} =10^4\  \cc$ and $10^6\  \cc$, respectively. 
In the high ionization state, such as $\log_{10} \ U = 0$, the $t^*(\civ)$ is less than 10 and 0.1 days for $n_{\rm H} =10^4\  \cc$ and $10^6\  \cc$, respectively. 

\subsection{Time-dependent photoionization simulations}

Although the response timescales predicted by equilibrium-state photoionization simulations in the previous section are consistent with the theoretical analysis, determining whether low- and high-ionization states exhibit positive or negative responses still requires verification through time-dependent simulations. Moreover, the response timescale given by Equation \ref{eq11}
is generally applicable only at the initial moment of a radiation change. Whether it can represent the characteristic timescale of the entire response process also needs to be examined 
through time-dependent simulations.

We use the \texttt{time} command in CLOUDY to perform time-dependent simulations.
As shown in Fig. \ref{fig6}, we select two ionization parameters, $\log_{10} U = -3$ and $\log_{10} U = 0$, to represent low- and high-ionization states, respectively. 
These two cases exhibit positive and negative responses, respectively. According to Equation \ref{eq11}, the response timescales for $\log_{10} U = -3$ and $\log_{10} U = 0$ are 
48.3 days and 5.5 days, respectively, indicating that the \civ\ concentration changes more rapidly at $\log_{10} U = 0$.
As shown in Figure 5, the results are fully consistent with the time-dependent simulations, which indicate that \civ\ concentration evolves more slowly at $\log_{10} U = -3$ 
than at $\log_{10} U = 0$. At the initial moment, the rate of change predicted by Equation \ref{eq11} coincides with the tangent to the simulation curve, 
further validating the theoretical analysis.
The response time scale of photoionized gases in this work ignores processes such as heating and cooling, and radiative transfer, e.g., \cite{sadaula2023}. Therefore, it is a simplification of the actual situation. But the predicted rate of change matches well with Cloudy simulations, indicating that processes such as heating and cooling, and radiative transfer are indeed unimportant.

The above analysis is based on the simplified assumption that the light curve behaves as a step function. To further examine the response of high- and low-ionization states in AGN light curves, we adopt the damped random walk (DRW) model \citep{kelly2009,kozlowski2009}. For a representative high-luminosity quasar, we assume a black hole mass of $\mbh = 10^9 \msun$, a characteristic timescale of $\tau = 30$ days, and a structure function defined as ${\rm SF}(\Delta t)={\rm SF_{\infty}}(1-e^{-|\Delta t|/\tau})^{1/2}$
with an asymptotic amplitude of ${\rm SF_{\infty}} = 0.2$ mag around $200$\,\AA~\citep{macleod2010}. 
For the ionizing photons, the relevant energy transitions correspond to 
\ciii~$\rightarrow$~\civ\ at $\sim 47.9$\,eV ($259$\,\AA) and \civ~$\rightarrow$~\cv\ at $\sim 64.5$\,eV ($192$\,\AA).
As shown in the top panel of Fig.~\ref{fig6}, we generate the light curve from the DRW model using the Python package \texttt{astroML} \citep{kelly2009}. 
The bottom panel of Fig.~\ref{fig6} illustrates the corresponding response. Consistent with the analysis above, in the low-ionization state ($\log_{10} U = -3$)
the $N_{\rm CIV}$ curve shows 
a positive response and appears smoother, indicating a longer response timescale. 
In contrast, in the high-ionization state ($\log_{10} U = 0$) the response is negative, and the curve is relatively less smooth, suggesting a shorter response timescale.

\section{Constraining BAL Gas Density through Asymmetric Response} 

In principle, an absorption line should vary from one observation to another only if  both the time over 
which the ionizing continuum varies and the time interval $\Delta T$ between the observations is longer than the $t^*$.
This asymmetry, i.e., the response time scale of high ionization states is shorter than that of low ionization states, suggests that in actual observations, 
the variations of absorption line in high ionization states (negative response) should 
are more easily detected than that in low ionization states (positive response). 
This happens to be consistent with actual observations, i.e., more over 70\% of BAL in quasar spectrum
exhibit a negative response \citep{wang2015, he2017}, based on large SDSS samples.

\begin{figure}[htb]
\centering
\includegraphics[width=0.5\textwidth]{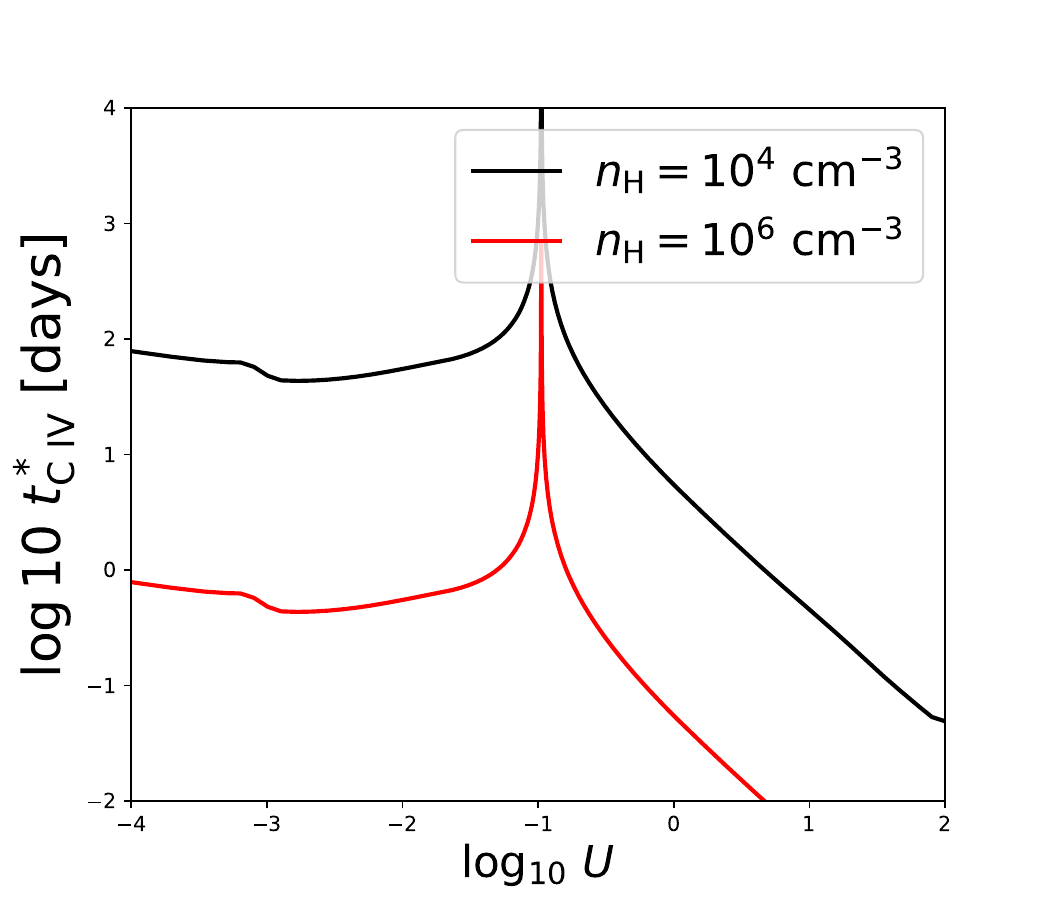}
\caption{The response timescale of \civ\ under different ionization parameters.
The black and red curves correspond to gas densities of $10^4\ \cc$ and $10^6\ \cc$, respectively.  
The timescale exhibits strong asymmetry: the response timescale of highly ionized states is significantly shorter than that of low-ionized states.  
If the observational time interval falls within 1--1000 days, this asymmetry would not be detectable for gas at a density of $10^6\ \cc$,  
but it would be observable for gas with a density of $10^4\ \cc$.
\label{fig5}}
\end{figure}

\begin{figure}[htb]
\centering
\includegraphics[width=0.5\textwidth]{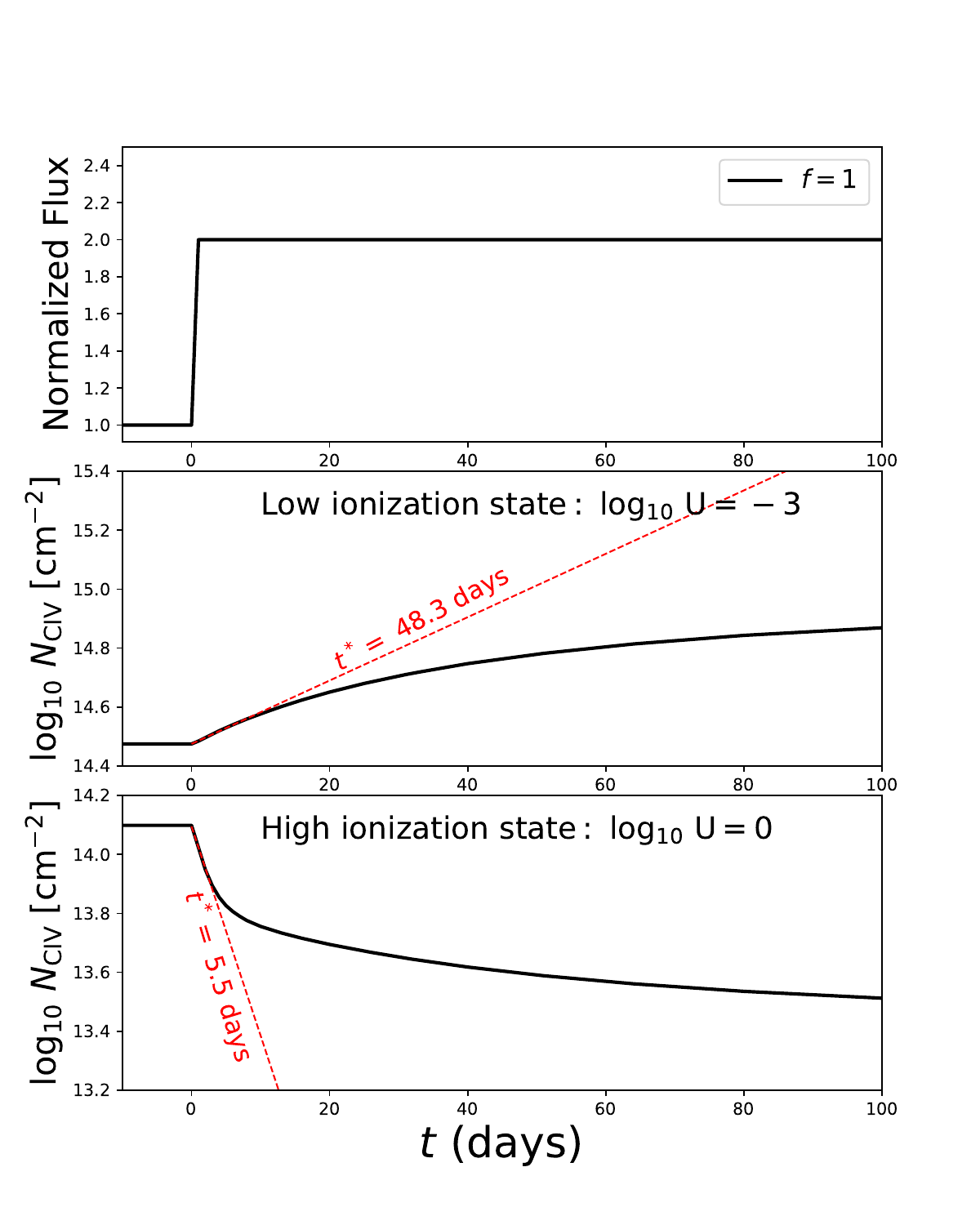}
\caption{Time-dependent photoionization simulations for the case of the light curve of the step function.
At $t = 0$, the radiation flux is suddenly increased from 1 to 2, corresponding to a change amplitude of $f = 1$.  
This leads to a positive response in the low-ionization state ($\log_{10} U = -3$) and a negative response in the high-ionization state ($\log_{10} U = 0$).  
The red dashed lines represent the response time scales at the initial time predicted by Equation \ref{eq11}.
The response timescales for $\log_{10} U = -3$ and $\log_{10} U = 0$ are 48.3 days and 5.5 days, respectively,  
indicating that the \civ\ concentration evolves more rapidly at higher ionization ($\log_{10} U = 0$).
\label{fig6}}
\end{figure}

The above negative responses exceeding 70\% can impose certain limitations on the density range of BAL gas.
As shown in Fig. \ref{fig5}, for gas with density $n_{\rm H} =10^4\  \cc$, the $t^*$ for low ionization states is about 100 days, 
while the $t^*$ for high ionization states is smaller than 100 days. 
Therefore, if the observation time interval is less than 100 days, we will only be able to detect responses in high ionized states, 
that is, only negative responses. If the gas density is $n_{\rm H} =10^6\  \cc$, the $t^*$ for low ionization states is about 1 days, 
while the $t^*$ for high ionization states is smaller than 1 day. At this point, both high and low ionization states can be detected in response.
Therefore, we can evaluate the range of BAL gas density based on the fraction of negative response.
In actual SDSS observations, the vast majority of observation time intervals  $\Delta T$ are greater than 1 day.
Here we make a rough assessment. We assume that the fraction of BAL gas with a density below $n_{\rm H} = 10^6\ \cc$ is $x$\%.  
If the number of intrinsic low-ionization and high-ionization states is the same, then the fraction of positive and negative responses observed for gases with a density above $n_{\rm H} = 10^6\ \cc$ is $0.5(1 - x\%)$.  
Assuming that all gases with a density below $n_{\rm H} = 10^6\ \cc$ exhibit negative responses (which is actually an overestimate, as $\Delta T > 1$ day),  
then the upper limit of the observed fraction of negative responses is $F$.  
Therefore, we may obtain a conservative estimate:
\[
0.5(1 - x\%) + x\% > 70\%, \quad \text{i.e.,} \quad x > 40\%.
\]
Thus, we conclude that at least 40\% of BAL gas must have a density below $n_{\rm H} = 10^6\ \cc$.

Interestingly, most measured BAL outflow gas densities reported in the literature fall below $10^6\ \cc$ \citep{korista2008,moe2009,dunn2010,borguet2013,lucy2014,chamberlain2015,
arav2018,leighly2018,hamann2019,he2019,he2022,xu2019,zhao2021,byun2022a,byun2022b,byun2023}.  
The typical gas density of an accretion disk wind \citep{murray1995,proga2000} is greater than $n_{\rm H} = 10^8\ \cc$.  
The fact that at least 40\% of observed BAL gas has densities below $10^6\ \cc$ suggests two possibilities:  
\begin{enumerate}
    \item The accretion disk wind has propagated to larger distances, leading to a decrease in density.  
    \item Some BAL outflows may originate from larger-scale structures, such as dusty torus winds \citep{konigl1994,scoville1995,elitzur2006,gallagher2015,he2022,ishibashi2024}, 
    whose spatial scale is over 100 times that of the accretion disk, resulting in significantly lower densities.
\end{enumerate}

For the sake of convenience in summarizing, we created a $t^*(\civ)$ map (Fig. \ref{fig8}) on the plane composed of $\log10 \ U$ and $\log10 \ n_e$. 
The three dashed lines in the figure represent contour lines for observation time intervals $\Delta T$ = 100, 10, and 1 day, respectively.
Below the respective dashed line, the observation time interval is smaller than the response timescale, i.e., $\Delta T < t^*_{\tiny \civ}$,
rendering it an unresponsive region. With observation time intervals of 100, 10, and 1 day, the lower density limits detectable for both low ionization and high 
ionization states are $10^{4}\  \cc$, $10^{5}\  \cc$ and $10^{6}\  \cc$, respectively. On the one hand, given a fixed observational time intervals, the lower 
the density of the gas, the more challenging it is to detect the response of the low ionized state. In other words, it becomes easier to detect the asymmetry of the response.
On the other hand, to detect the response asymmetry of higher density gases, a shorter observation time interval is required.
It should be pointed out that this is just a simple quantitative analysis, but it at least indicates the existence of low-density gas in BAL gas that is very far away from the SMBH. Some studies have revealed that BAL gas is likely low-density and can exist at scales beyond dust torus, e.g., \cite{zhang2014,he2019,he2022,naddaf2023, ishibashi2024}, which is consistent with our results.

\begin{figure}[htb]
\centering
\includegraphics[width=0.5\textwidth]{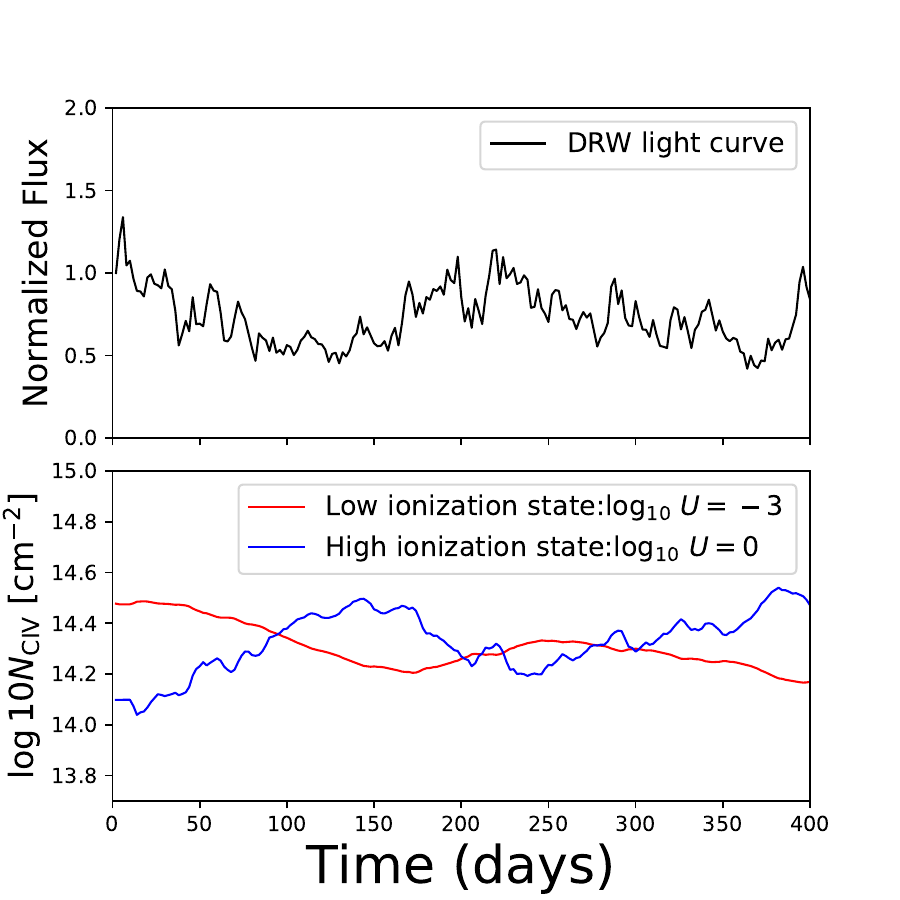}
\caption{Time-dependent photoionization simulations for the case of the DRW light curve.
Top panel: the DRW light curve.  
Bottom panel: the gas response at different ionization states, showing a positive response in the low-ionization state ($\log_{10} U = -3$) and 
a negative response in the high-ionization state ($\log_{10} U = 0$).  
In addition, in the low-ionization state the $N_{\rm CIV}$ curve appears smoother, indicating that its response timescale is longer than that in the high-ionization state.
\label{fig7}}
\end{figure}

\begin{figure}[htb]
\centering
\includegraphics[width=0.5\textwidth]{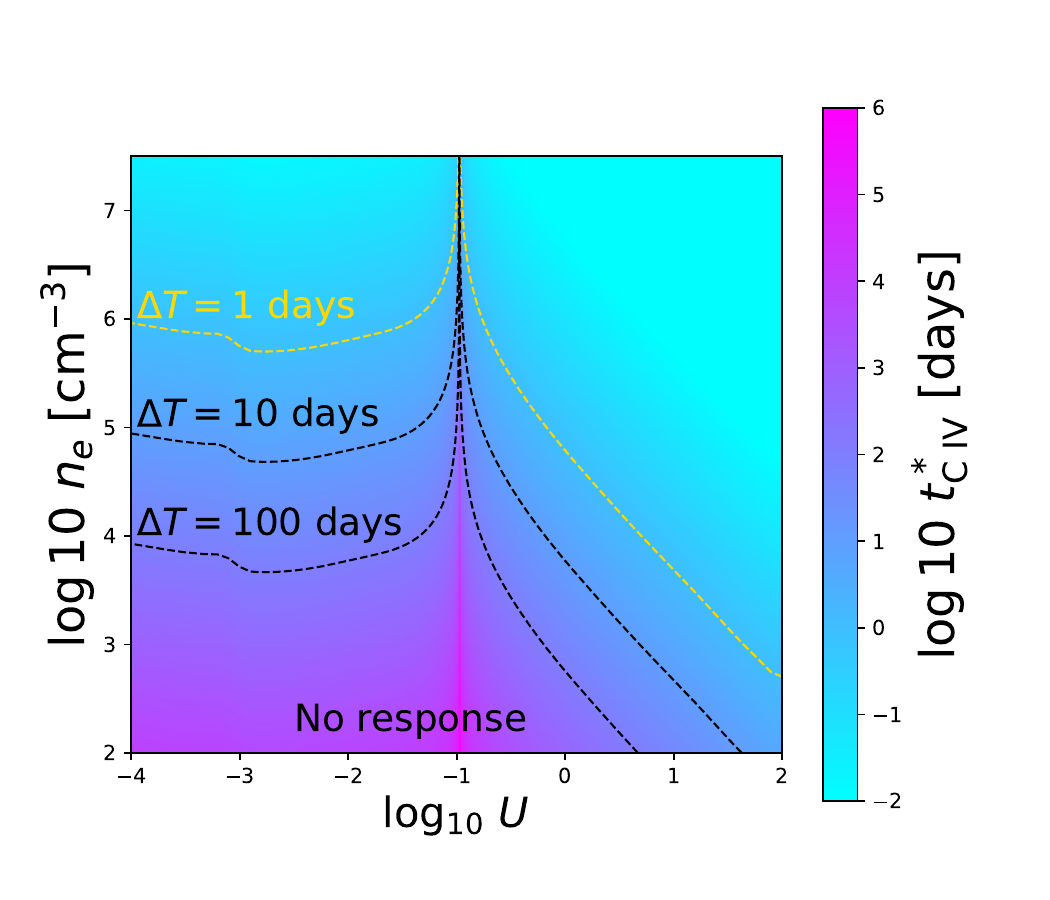}
\caption{The map of \civ\ response timescales on the plane of ionization parameter and gas density.
The dashed lines represent contours corresponding to observation intervals $\Delta T$ of 1, 10, and 100 days.  
The golden dashed line marks 1 day, the typical minimum time interval of SDSS observations.  
Below each contour, the observation interval is shorter than the response timescale,  
indicating a non-responsive region where ion concentration changes cannot be detected.
\label{fig8}}
\end{figure}

\section{Summary}
\label{sec:summary}

In this work, we report the discovery of an asymmetric response in photoionized gas. In low ionization states, a positive correlation—termed a \emph{positive response}—exists between absorption lines and the ionizing source. In contrast, highly ionized states exhibit an inverse relationship. Large-scale statistical analyses from the Sloan Digital Sky Survey (SDSS) reveal a striking asymmetry: over 70\% of BAL (Broad Absorption Line) gases in quasar host galaxies display a negative response to variations in quasar radiation.

Our key findings are as follows:

\begin{enumerate}
    \item Through analytical derivations and photoionization simulations of \civ, we demonstrate that the response of gas to radiation is inherently asymmetric between low and high ionization states.
    
    In sufficiently low ionization conditions, i.e., $n_{i+1}/n_i \ll \alpha_{i-1}/\alpha_i$ or $I_i \ll n_e \alpha_{i-1}$, the response timescale is given by
    \[
    t_i^* = \frac{1}{f n_e \alpha_{i-1}}.
    \]
    In contrast, for highly ionized states where $n_{i+1}/n_i \gg \alpha_{i-1}/\alpha_i$ or $I_i \gg n_e \alpha_{i-1}$, the response timescale becomes
    \[
    t_i^* = -\frac{1}{f \alpha_i n_e \frac{n_{i+1}}{n_i}} = -\frac{1}{f I_i} = -\frac{r^2}{f \Gamma_i}.
    \]
    Thus, in low ionization states, the response is governed by the recombination coefficient, while in high ionization states, it is governed by the ionization rate. This leads to shorter response timescales and a greater likelihood of observing negative responses in high ionization states.

    \item This asymmetric effect provides a natural explanation for why over 70\% of BALs exhibit negative responses. Based on this, we further make a conservative estimate that at least 40\% of BAL gas must have a density below $n_{\rm H} = 10^6\ \cc$, consistent with the fact that most measured BAL gas densities fall below this threshold. In contrast, the typical gas density of an accretion disk wind is generally higher than $n_{\rm H} = 10^8\ \cc$. This discrepancy implies that either the accretion disk wind has expanded to larger scales, resulting in a lower density, or that BAL outflows may originate from more extended regions, such as dust torus winds.
\end{enumerate}

While the asymmetry in photoionization gas response offers a reasonable explanation for the observed excess of negative responses, it may not be the sole cause. To fully understand the prevalence of negative responses in absorption-line systems, further studies are needed—especially to determine whether this asymmetry persists for ions with higher ionization potentials (e.g., \ovi\ at 138.1 eV) or even for highly ionized iron species ($>1$ keV). Another promising avenue for testing this effect is the study of absorption lines originating from stellar winds \citep{lamers1999} in stellar spectra.

\begin{acknowledgements}
      Zhicheng He is supported by the USTC Research Launch Project KY2030000187 and the National Natural Science Foundation of China (nos. 12222304, 12192220, and 12192221).
\end{acknowledgements}

\bibliographystyle{aa}

\end{document}